\documentclass[preprint]{ptephy_v1}

\preprintnumber{
RIKEN-NC-NP-181, RIKEN-QHP-331
} 

\usepackage{amsmath,amssymb}
\usepackage{graphics} 
\usepackage[dvipdfmx]{hyperref}

\newcommand{\vslash}{{v\hspace{-5.4pt}/}}

\begin{document}

\title{Mesic nuclei with a heavy antiquark}


\author{Yasuhiro Yamaguchi\email{yasuhiro.yamaguchi@ge.infn.it}}
\affil{Istituto Nazionale di Fisica Nucleare (INFN), Sezione di
Genova, via Dodecaneso 33, 16146 Genova, Italy, \\
Theoretical Research Division, 
Nishina Center, RIKEN,
Wako, Saitama 351-0198, Japan\\
}
\author{\\ Shigehiro Yasui}
\affil{Department of Physics, Tokyo Institute of Technology,
Tokyo 152-8551, Japan}




 \begin{abstract}%
   The binding 
   of a hadron and a nucleus is a topic of great
   interest 
  for investigating
   the hadron properties.
   In the heavy-flavor region,
   attraction between a $P(=\bar{D},B)$ meson and a nucleon $N$ can appear,
   where 
   $PN-P^\ast N$ mixing plays an important role in relation to 
   the heavy-quark spin symmetry.
   The attraction can produce exotic heavy mesic nuclei
   that are stable against strong decay.
   We study an exotic system where the $\bar{D}$ ($B$) meson and nucleus are bound.
   The meson-nucleus interaction is
   given by a folding potential with single-channel $PN$ interaction and the nucleon number distribution
   function.
   By solving the Schr\"odinger equations of
   the heavy meson and the nucleus,
  we obtain several bound and resonant states for nucleon number
  $A=16,\dots,208$.
   The results indicate the possible existence of exotic mesic
   nuclei with a heavy antiquark.
 \end{abstract}

\subjectindex{D06, D15}


\maketitle

\section{Introduction}

Multiflavor nuclei are interesting objects for studying unconventional states of matter in the hadron and nuclear physics.
As a first step 
towards extending
the quark flavor, strangeness nuclei have been extensively studied both in experiments and in theories.
As a new direction, charm and bottom are new flavors whose properties in nuclei should be different from those of strangeness nuclei~\cite{Hosaka:2016ypm}.
Charm/bottom nuclei, as well as strangeness nuclei, are important for studying 
(i) the heavy hadron and nucleon interaction, (ii) the properties of heavy hadrons in a nuclear medium, and 
(iii) the impurity effect for nuclear properties.
These are related to the flavor symmetry of the interhadron interaction, the partial restoration of the broken chiral symmetry, and so on, as fundamental problems in quantum chromodynamics (QCD).
In theoretical studies, 
there 
has been a lot of research into
heavy baryons, heavy mesons and quarkonia in nuclear systems 
(see Ref.~\cite{Hosaka:2016ypm} and references therein).
Recently, few-body calculations have also been performed~\cite{Bayar:2012dd,Yokota:2013sfa,Yamaguchi:2013hsa,Maeda:2015hxa}.
In the present study, we focus on the properties of the $\bar{D}$ ($B$) mesons with 
a quark content $\bar{Q}q$ with a heavy antiquark $\bar{Q}$ and a light quark $q$, which can be bound in finite-size atomic nuclei.

The $\bar{D}$ ($B$) meson in nuclear systems is a simple system, 
because there is no annihilation channel by $\bar{q}q$, 
in contrast to the case of its antiparticle $D$ ($\bar{B}$) in nuclear medium.
They have both been studied in many theoretical works:
the quark-meson coupling model~\cite{Tsushima:1998ru,Sibirtsev:1999js}, 
the mean-field
approach~\cite{Mishra:2003se,Mishra:2008cd,Kumar:2010gb,Kumar:2011ff},
the flavor SU(4)
symmetry~\cite{Lutz:2005vx,Mizutani:2006vq,Tolos:2007vh,JimenezTejero:2011fc},
the flavor-spin SU(8) symmetry~\cite{GarciaRecio:2011xt}, the
pion-exchange interaction~\cite{Yasui:2012rw,Yasui:2013iga}, the QCD sum
rules~\cite{Hayashigaki:2000es,Azizi:2014bba,Wang:2015uya,Hilger:2008jg,Hilger:2010zb,Suzuki:2015est},
and the Nambu--Jona-Lasinio model~\cite{Blaschke:2011yv}.
As advanced topics, there are studies on chiral symmetry
breaking~\cite{Sasaki:2014asa,Sasaki:2014wma,Suenaga:2014sga,Suenaga:2015daa},
the Kondo
effect~\cite{Yasui:2013xr,Yasui:2016ngy,Yasui:2016hlz}\footnote{The
Kondo effect is also considered in quark matter, where the color degrees
of freedom play an essential
role~\cite{Hattori:2015hka,Ozaki:2015sya,Yasui:2016svc,Yasui:2016yet,Kanazawa:2016ihl}.},
and the spin-isospin correlated nuclear matter~\cite{Suenaga:2014dia}.
Few-body calculations of $\bar{D}NN-\bar{D}^{\ast}NN$ ($BNN-B^{\ast}NN$) have also been performed~\cite{Yamaguchi:2013hsa}.
The interaction between a $\bar{D}$ ($B$) meson and a nucleon can be provided by the meson exchange interaction.
At short distances, it is supplied by the exchange of heavy mesons (e.g., scalar mesons and vector mesons) and by the direct exchange of quarks~\cite{Hofmann:2005sw,GarciaRecio:2008dp,Haidenbauer:2007jq,Carames:2012bd,Carames:2016qhr}.
At long distances, 
pion exchange can occur, because the
pion is the lightest meson as the Nambu-Goldstone bosons generated by
the dynamical breaking of chiral symmetry in
vacuum~\cite{Cohen:2005bx,Haidenbauer:2007jq,Yasui:2009bz,Yamaguchi:2011xb,Yamaguchi:2011qw,Carames:2012bd,Carames:2016qhr}.

The symmetry that governs the $\bar{D}$ ($B$) meson dynamics is given by 
the heavy-quark (spin) symmetry (HQS) for the heavy-antiquark component and chiral symmetry 
for the light-quark component~\cite{Neubert:1993mb,Casalbuoni:1996pg,Manohar:2000dt}.
One of the most important properties of the HQS is that the spin degree of freedom of a heavy (anti)quark is decoupled from the spatial rotation.
Then, the heavy-quark spin is independent of the light spin $j$, which is carried by light quarks and gluons, as a sum of angular momenta and intrinsic spins.
As a consequence, there exist two different types of heavy hadron states: the HQS singlet ($j=0$) and the HQS doublet ($j\ge 1/2$)~\cite{Neubert:1993mb,Casalbuoni:1996pg,Manohar:2000dt,Yasui:2013vca,Yamaguchi:2014era}.
A $\bar{D}$ ($B$) meson is regarded approximately as an HQS doublet whose pairs are given by a $\bar{D}^{\ast}$ ($B^{\ast}$) meson, because the mass difference is of the order of $1/m_{c}$ ($1/m_{b}$) with $m_{c}$ ($m_{b}$) being the charm (bottom) quark mass.
This is a good approximation because the charm (bottom) quark mass is larger than the typical low-energy scale of the hadron dynamics, say a few hundred MeV, and they can be regarded as being sufficiently heavy.
One of the interesting results of the HQS is that the mixing of a $\bar{D}$ ($B$) meson 
and a $\bar{D}^{\ast}$ ($B^{\ast}$) meson is realized in nuclear medium via the interaction process $\bar{D}N \leftrightarrow \bar{D}^{\ast}N$ ($BN \leftrightarrow B^{\ast}N$).
This mixing leads to an attraction of the heavy-light hadrons in nuclear matter~\footnote{This is a two-body mixing for spin degrees of freedom. An analogous process is seen in hypernuclei, i.e., $\Lambda N \leftrightarrow \Sigma N$ for isospin mixing. 
When the spin-isospin correlation exists in nuclear matter, one-body mixing by $\bar{D} \leftrightarrow \bar{D}^{\ast}$ ($B \leftrightarrow B^{\ast}$) can exist~\cite{Suenaga:2014dia}.}.

The current status of theoretical studies, however, is that 
there are many open problems about the properties of the $\bar{D}$ ($B$) meson and the $\bar{D}^{\ast}$ ($B^{\ast}$) meson in nuclear systems.
For example, the value of the binding energy of the $\bar{D}$ ($B$) meson and/or the $\bar{D}^{\ast}$ ($B^{\ast}$) meson in nuclear matter with infinite volume is not yet convergent.
The values are highly scattered from a few MeV to a hundred MeV depending on the model used in the analysis.
Hence more theoretical effort is needed to understand heavy hadrons in nuclear medium.

The purpose of the present study is to investigate the bound and/or resonant states of 
the $\bar{D}$ ($B$) meson in atomic nuclei with finite nucleon numbers.
As an interaction between a $\bar{D}$ ($B$) meson and a nucleon, we adopt the pion exchange potential at long distances and 
the vector-meson exchange potential at short distances.
The existence of the pion exchange interaction is an important result of the HQS, 
 because the mixing process $\bar{D}N \leftrightarrow \bar{D}^{\ast}N$ ($BN \leftrightarrow B^{\ast}N$) 
 plays an important role~\cite{Yasui:2009bz,Yamaguchi:2011xb,Yamaguchi:2011qw}.
At first sight, there seems to be no $\bar{D}\bar{D}\pi$ ($BB\pi$) vertex 
by parity conservation, and hence there should be no pion exchange interaction for $\bar{D}N$ ($BN$).
The pion exchange interaction for $\bar{D}N$ ($BN$) is indeed induced by the two-step process $\bar{D}N \rightarrow \bar{D}^{\ast}N \rightarrow \bar{D}N$ ($BN \rightarrow B^{\ast}N \rightarrow BN$).
This mixing is available 
when the mass of a $\bar{D}^{\ast}$ ($B^{\ast}$) meson is sufficiently close to the mass of a $\bar{D}$ ($B$) meson, 
as expected from  the (approximate) mass degeneracy in the heavy-quark limit.
In this framework, we eliminate the dynamical degrees of freedom by a $\bar{D}^{\ast}$ ($B^{\ast}$) meson in a systematic way, and obtain the effective interaction between a $\bar{D}$ ($B$) meson and a nucleon.
We then apply the obtained $\bar{D}N$ ($BN$) potential to the calculation of the energy levels of a $\bar{D}$ ($B$) meson in finite-size atomic nuclei.

This paper is organized as follows.
In Sect.~\ref{sec:potential}, we formulate
the meson exchange potential between a
heavy meson and a nucleon, perform the
projection of the meson exchange potential, and give the meson-nucleus folding potential. 
In Sect.~\ref{sec:result}, we show the numerical results of the bound and resonant states.
The last section is devoted to a summary.

\section{Interactions between the heavy meson and the nucleus}
\label{sec:potential}

The heavy meson in a nucleus is regarded as a two-body system of the
meson and the nucleus.
The meson-nucleus interaction is described as the folding potential, which can be
obtained by the $P$ meson-nucleon ($PN$) interaction and the nucleon number
distribution function in the nucleus.
Hereafter, we will use the notation $P$ to stand for a $\bar{D}$ meson or a $B$ meson.
We will also use the notation $P^{\ast}$ for a $\bar{D}^{\ast}$ meson or a $B^{\ast}$ meson.  
As for the $PN$ potential, we employ the meson exchange
potential of the coupled-channel $PN-P^{\ast}N$, 
as discussed in Refs.~\cite{Yasui:2009bz,Yamaguchi:2011xb,Yamaguchi:2011qw}.
The most simple $P^{(\ast)}N$ system is the $S$-wave state with the
total spin $J=1/2$.
In Refs.~\cite{Yasui:2009bz,Yamaguchi:2011xb,Yamaguchi:2011qw}, we found 
one bound state for $I=0$, and no bound state for
$I=1$. 
Therefore, the $I(J^P)=0(1/2^-)$ state
is the most important one in the mesic nuclear system.

The $I(J^P)=0(1/2^-)$ state has three channels, namely 
$PN(^{2}S_{1/2})$, $P^{\ast}N(^{2}S_{1/2})$ and
$P^{\ast}N(^{4}D_{1/2})$, where the notation $^{2S+1}L_{J}$ is used to stand for the intrinsic total spin $S$, the angular momentum $L$ and the total spin $J$.
This potential is given by the $3\times 3$ matrix form.
We will project this potential onto the $PN(^2S_{1/2})$
channel,
and use the single-channel $PN$ potential to construct the
folding potential.

\subsection{Meson exchange potentials of $P^{(\ast)}$ meson and nucleon $N$}
\label{sec:PNpotential}

First of all, we introduce the meson exchange potential between a $P^{(\ast)}$ meson and a nucleon
$N$.
This potential is obtained by the effective Lagrangians of
heavy mesons, light mesons, and nucleons,
as discussed in Refs.~\cite{Yasui:2009bz,Yamaguchi:2011xb,Yamaguchi:2011qw}.

The Lagrangians of heavy mesons and light mesons (pion and vector mesons),
satisfying the heavy quark and chiral symmetries~\cite{Wise:1992hn,Falk:1992cx,Yan:1992gz} 
(see also Refs.~\cite{Neubert:1993mb,Casalbuoni:1996pg,Manohar:2000dt}), are employed in this
study.
Their interaction forms are given by
\begin{align}
 &{\cal L}_{\pi HH} = ig_\pi{\rm Tr}\left[H_b \gamma_\mu\gamma_5
 A^{\mu}_{ba}\bar{H}_a\right],
 \label{eq:LpiHH}
 \\
 &{\cal L}_{v HH} = -i\beta{\rm Tr}\left[H_b v^\mu
 \left(\rho_\mu\right)_{ba}\bar{H}_a\right]
 +i\lambda{\rm Tr}\left[H_b\sigma^{\mu\nu}F_{\mu\nu}(\rho)_{ba}\bar{H}_a\right].
 \label{eq:LvHH}
\end{align}
The subscripts $\pi$ and $v$ stand for the pion and vector mesons
($\rho$ and $\omega$), respectively.
$v^\mu$ is the four-velocity of a heavy quark ($v^{\mu}v_{\mu}=1$ and $v^{0}>0$).
The subscripts $a,b$ are for light flavor ($u,d$).
The heavy-meson fields $H_a$ are given by
\begin{align}
 &H_a=\frac{1+\vslash}{2}\left[P^\ast_{a\mu}\gamma^\mu-P_a
 \gamma_5\right],
 \quad \bar{H}_a=\gamma_0H^\dagger_a \gamma_0 ,
\end{align}
where $P^{\ast}_{a\mu}$ and $P_{a}$ are the fields of $P^{\ast}$ and $P$ mesons, respectively, with light flavor $a$.
$A^\mu$ is the axial current of a pion, expressed as
\begin{align}
 &A^\mu=\frac{1}{2}\left(\xi^\dagger\partial^\mu\xi-\xi\partial^\mu\xi^\dagger\right),
\end{align}
with $\xi=\exp(i\hat{\pi}/f_\pi)$, the pion decay constant $f_\pi=132$
MeV, and the pion field
\begin{align}
 &\hat{\pi}=\left(
 \begin{array}{cc}
  \frac{\pi^0}{\sqrt{2}}
   &\pi^+ \\
  \pi^-&-\frac{\pi^0}{\sqrt{2}}
   \\
 \end{array}
 \right).
\end{align}
The vector-meson field $\rho_\mu$ (for $\rho$ and $\omega$ mesons) and the field tensor
$F_{\mu\nu}(\rho)$ are given by
\begin{align}
 &\rho_\mu=i\frac{g_V}{\sqrt{2}}\hat{\rho}_\mu,\\
 &\hat{\rho}_\mu=\left(
 \begin{array}{cc}
  \frac{\rho^0}{\sqrt{2}}
   +\frac{\omega}{\sqrt{2}}
  &\rho^+ \\
  \rho^-&
   \frac{-\rho^0}{\sqrt{2}}
   +\frac{\omega}{\sqrt{2}}
 \end{array}
 \right)_\mu,\\
 &F_{\mu\nu}(\rho)=\partial_\mu\rho_\nu-\partial_\nu\rho_\mu +
 \left[\rho_\mu, \rho_\nu\right],
\end{align}
where $g_V=m_V/f_\pi$ is the gauge-coupling constant of the hidden local
symmetry~\cite{Bando:1987br}.

The coupling constant $g_\pi$ in Eq.~\eqref{eq:LpiHH}
is determined by the $D^\ast\rightarrow D\pi$
decay~\cite{Casalbuoni:1996pg,Manohar:2000dt,Olive:2016xmw}.
For the vector mesons in Eq.~\eqref{eq:LvHH}, $\beta$ and $\lambda$
are fixed by radial decays of $D^\ast$ and semileptonic decays of $B$
with vector-meson dominance~\cite{Isola:2003fh}.
The values of the coupling constants are summarized in Table~\ref{table:couplings}.

\begin{table}[tbp]
 \caption{Masses of light mesons $\alpha(=\pi,\rho,\omega)$, $m_{\alpha}$, and coupling constants of the
 Lagrangians.
 $m_{\alpha}$ and $\lambda$ are given in units of MeV and GeV$^{-1}$,
 respectively.
 The other parameters are dimensionless constants.}
 \label{table:couplings}
   \begin{center}
    \begin{tabular}{lcccccc}
     \hline\hline
     $\alpha$& $m_\alpha$ [MeV]& $g_{\pi}$& $\beta$& $\lambda$ [GeV$^{-1}$]&
     $g^2_{\alpha NN}/4\pi$& $\kappa$\\ \hline
     $\pi$&132.7 &0.59 &--- &--- &13.6 &--- \\
     $\rho$&769.9 &--- &0.9 &0.56 &0.84 &6.1 \\
     $\omega$&781.94 &--- &0.9 &0.56 &20.0 &0.0 \\
     \hline\hline
    \end{tabular}
   \end{center}
\end{table}

The effective Lagrangians for the interaction vertices of nucleons and light mesons are
given by
 \begin{align}
  &{\cal L}_{\pi NN}=\sqrt{2}ig_{\pi NN}\bar{N}\gamma_5 \hat{\pi}N ,
  \label{eq:LpiNN} \\
  &{\cal L}_{v NN}=\sqrt{2}g_{v NN}\left[\bar{N}\gamma_\mu \hat{\rho}^\mu  N
  +\frac{\kappa}{2m_N}\bar{N}\sigma_{\mu\nu}\partial^\nu \hat{\rho}^\mu  N\right] ,
    \label{eq:LvNN}
 \end{align}
as shown in Refs.~\cite{Machleidt:1987hj,Machleidt:2000ge}.
 The nucleon field is expressed by $N=(p,n)^t$.
 The coupling constants, $g_{\pi NN}$, $g_{vNN}$ and
 $\kappa$, are summarized in Table~\ref{table:couplings}.

 In order to parameterize the internal structure of hadrons,
 the dipole form factor is introduced at each vertex:
 \begin{align}
  &F_{\alpha}(\Lambda, \vec{q}\,)=
  \frac{\Lambda^2-m^2_\alpha}{\Lambda^2+\left|\vec{q}\,\right|^2},
  \label{eq:formfactor}
 \end{align}
 with the mass $m_\alpha$ and the three-momentum $\vec{q}$ of incoming
 light mesons $\alpha=\pi,\rho,\omega $.
 As discussed in Refs.~\cite{Yasui:2009bz,Yamaguchi:2011xb,Yamaguchi:2011qw},
 the cutoff parameter $\Lambda=\Lambda_N$ for 
 the nucleon
 vertex is
 determined to reproduce the binding energy of a deuteron.
 $\Lambda=\Lambda_P$ for 
 the heavy-meson vertex is fixed
 by the ratio of the sizes of the pseudoscalar (vector) meson $P^{(\ast)}$ and the
 nucleon.
 We use $\Lambda_N=846$ MeV for the nucleon, $\Lambda_{D^{(\ast)}}=1.35\Lambda_N$
 for the $\bar{D}^{(\ast)}$ meson and
 $\Lambda_{B^{(\ast)}}=1.29\Lambda_N$ for the $B^{(\ast)}$ meson in Refs.~\cite{Yasui:2009bz,Yamaguchi:2011xb,Yamaguchi:2011qw}.

From the effective Lagrangians in Eqs.~\eqref{eq:LpiHH}-\eqref{eq:LvNN}
and the form factor in Eq.~\eqref{eq:formfactor},
the meson exchange potential between a $P^{(\ast)}$ meson and a nucleon
$N$ is obtained.
The one-pion exchange potential (OPEP) is expressed by
\begin{align}
 &V^{\pi}_{PN-P^\ast N}(r)=-\frac{g_\pi g_{\pi NN}}{\sqrt{2}m_Nf_\pi} \frac{1}{3}
 \left[\vec{\varepsilon}\,^\dagger\cdot\vec{\sigma}
 C(r;m_\pi)+S_{\varepsilon}T(r;m_\pi)\right]\vec{\tau}_P\cdot\vec{\tau}_N, \\
 &V^{\pi}_{P^\ast N-P^\ast N}(r)=\frac{g_\pi g_{\pi NN}}{\sqrt{2}m_Nf_\pi} \frac{1}{3}
 \left[\vec{S}\cdot\vec{\sigma}
 C(r;m_\pi)+S_{S}T(r;m_\pi)\right]\vec{\tau}_P\cdot\vec{\tau}_N,
\end{align}
with $r$ being the distance between $P^{(\ast)}$ and $N$.
$m_N=940$ MeV is the mass of a nucleon.
$\vec{\varepsilon}$
($\vec{\varepsilon}\,^\dagger$) is the polarization
vector of the incoming (outgoing) heavy vector meson $P^\ast$.
$\vec{S}$ is the spin-one operator given by $\vec{S}=i\vec{\varepsilon}\,^\dagger\times\vec{\varepsilon}$.
$\vec{\sigma}$ ($\vec{\tau}$) are the Pauli matrices for the (iso)spin.
$S_{\cal O}(\hat{r})$ is the tensor operator expressed by
$S_{\cal O}(\hat{r})=3(\vec{\cal O}\cdot\hat{r})(\vec{\sigma}\cdot\hat{r})-\vec{\cal O}\cdot\vec{\sigma}$
for ${\cal O}=\varepsilon,S$.
The functions $C(r;m_{\alpha})$ and $T(r;m_{\alpha})$ are given by
\begin{align}
 &C(r;m_{\alpha})=\int\frac{d^3q}{(2\pi)^3}\frac{m_{\alpha}^2}{\vec{q}\,^2+m_{\alpha}^2}e^{i\vec{q}\cdot\vec{r}}
 F(\Lambda_P,\vec{q}\,)F(\Lambda_N,\vec{q}\,), \\
 &S_{{\cal O}}(\hat{r})T(r;m_{\alpha})=\int\frac{d^3q}{(2\pi)^3}\frac{-\vec{q}\,^2}{\vec{q}\,^2+m_{\alpha}^2}
 S_{{\cal O}}(\hat{q})e^{i\vec{q}\cdot\vec{r}}
 F(\Lambda_P,\vec{q}\,)F(\Lambda_N,\vec{q}\,).
\end{align}
Similarly, the vector-meson exchange potentials are given by
\begin{align}
 &V^{v}_{PN-PN}(r)=\frac{\beta
 g_Vg_{vNN}}{\sqrt{2}m^2_v}C(r;m_v)\vec{\tau}_P\cdot\vec{\tau}_N,\\  
 &V^{v}_{PN-P^\ast N}(r)=
 \frac{g_Vg_{vNN}\lambda(1+\kappa)}{\sqrt{2}m_N}\frac{1}{3}
 \left[-2\vec{\varepsilon}\cdot\vec{\sigma}C(r;m_v)+S_{\varepsilon}T(r;m_v)\right]\vec{\tau}_P\cdot\vec{\tau}_N,\\
 &V^{v}_{P^\ast N-P^\ast N}(r)=\frac{\beta
 g_Vg_{vNN}}{\sqrt{2}m^2_v}C(r;m_v)\vec{\tau}_P\cdot\vec{\tau}_N \notag\\
 &\hspace{2.35cm}+\frac{g_Vg_{vNN}\lambda(1+\kappa)}{\sqrt{2}m_N}\frac{1}{3}
 \left[2\vec{S}\cdot\vec{\sigma}C(r;m_v)-S_{S}T(r;m_v)\right]\vec{\tau}_P\cdot\vec{\tau}_N.
\end{align}
The $P^{(\ast)}N$ potential in the $I(J^P)=0(1/2^-)$ state is presented by the $3\times 3$ matrix form.
The Schr\"odinger equation is given by
\begin{eqnarray}
\left(
\begin{array}{ccc}
 K_{0} +V_{11} & V_{12} & V_{13}  \\
 V_{12} & K_{0} +V_{22} & V_{23}  \\
 V_{13} & V_{23} & K_{2}+V_{33}   
\end{array}
\right)
\left(
\begin{array}{c}
 \psi_{1}^{E}  \\
 \psi_{2}^{E}  \\
 \psi_{3}^{E}  \\ 
\end{array}
\right)
=
E
\left(
\begin{array}{c}
 \psi_{1}^{E}  \\
 \psi_{2}^{E}  \\
 \psi_{3}^{E}  \\ 
\end{array}
\right),
\label{eq:Hamiltonian}
\end{eqnarray}
with the kinetic term $K_{\ell}$ with angular momentum $\ell$.
The channels $i=1,2,3$ correspond to $PN(^2S_{1/2}),P^\ast N(^2S_{1/2})$
and $P^\ast N(^4D_{1/2})$, respectively.
The potential $V_{ij}(r)$
is the sum of the $\pi$, $\rho$, $\omega$ potentials.
The explicit forms of the matrix elements as well as the kinetic terms are given in the appendix in Ref.~\cite{Yamaguchi:2011xb}.
As shown in Ref.~\cite{Yamaguchi:2011xb}, from Eq.~(\ref{eq:Hamiltonian}), we obtain the eigenvalues $E=2.14$ MeV for the $\bar{D}N$ bound state and $ 23.04$ MeV for the $BN$ bound state.

Let us comment on the interaction employed in this study.
We obtain cutoff parameters that are close to the vector-meson masses, and hence
the form factor~\eqref{eq:formfactor}
 would suppress the contribution from the short-range interaction as the vector-meson exchanges.
However, we consider that this short-range interaction includes not only the vector-meson exchange contribution, 
but also others: the scalar-meson exchange, the $\eta$ exchange, the $\pi\pi$ exchange, etc.
In Refs.~\cite{Yasui:2009bz,Yamaguchi:2011xb},
we employ the $\pi\rho\omega$ potentials as the nucleon-nucleon interaction, while
various short-range interactions can work, as shown in Refs.~\cite{Machleidt:1987hj,Machleidt:2000ge}.
Then, the cutoffs of the $\pi\rho\omega$ potentials can be determined as $\Lambda_N=846$ MeV to reproduce the deuteron properties.
In this parameter fitting, the small contribution from the vector-meson exchanges 
is interpreted as the cancellation of the various short-range interactions with each other.
In particular, the scalar-meson exchange provides a strong attraction, 
and it can suppress the repulsion of the vector-meson exchange potential, 
as seen in the Bonn potential~\cite{Machleidt:1987hj,Machleidt:2000ge}.
The potential employed in this study is also considered to include various meson exchanges,
because the cutoffs fixed in the deuteron are used to determine the cutoffs at vertices of the heavy meson and nucleon.
Therefore, our vector-meson exchange potential is not pure, 
but we regard it as a potential including various short-range contributions.

 \subsection{Folding potential}

We focus on the $I(J^{P})=0(1/2^{-})$ state as the most attractive one, 
with three channels $PN(^{2}S_{1/2})$, $P^{\ast}N(^{2}S_{1/2})$, and $P^{\ast}N(^{4}D_{1/2})$.
 To construct the folding potential between the $P$ meson and the nucleus, 
we project the $PN-P^{\ast}N$ coupled-channel potential 
 onto the $PN$ single-channel potential.

\begin{figure}[t]
  \begin{center}
   \includegraphics[width=0.7\textwidth,clip]{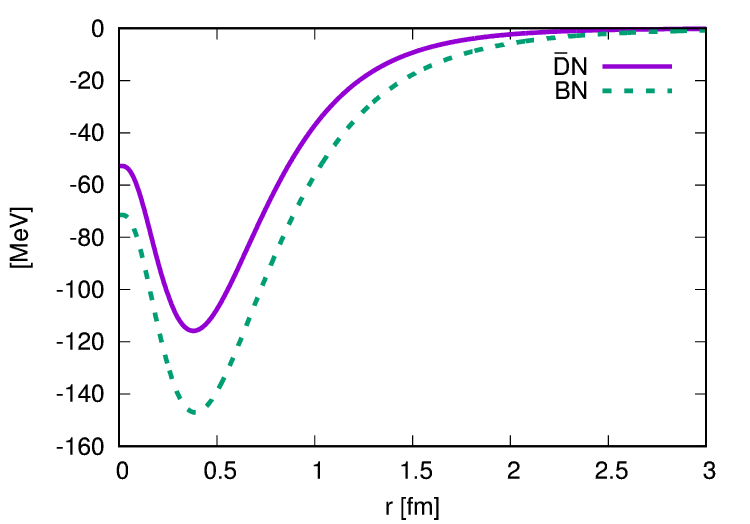}
  \end{center}
 \caption{The single-channel $\bar{D}N$ and $BN$ potentials.
 The $\bar{D}N$ ($BN$) potential is plotted by the solid (dashed) line.}
 \label{fig:Projectedpote}
\end{figure}

The projection onto the $PN(^2S_{1/2})$ channel is performed by
\begin{align}
 V_{\rm pro}(r)=V_{11}(r)+V_{12}(r)\frac{\psi^E_{2}}{\psi^E_{1}}+V_{13}(r)\frac{\psi^E_{3}}{\psi^E_{1}},
 \label{v_pro}
\end{align}
with $\psi^E_{i}$ ($i=1,2,3$) in the coupled-channel Schr\"odinger
equation of the $P^{(\ast)}N$ two-body system (\ref{eq:Hamiltonian}).
Here $E$ is the binding energy for $PN$.
Then, the single-channel $PN$ potential is obtained as shown in Fig.~\ref{fig:Projectedpote}.
We notice that $V_{\rm pro}(r)$ is the energy-dependent potential by its definition.
However, in what follows, 
we assume that $V_{\rm pro}(r)$ 
does not qualitatively change
so much for different $E$ at normal nuclear matter density, and try to investigate the energy levels of a $\bar{D}$ ($B$) meson in atomic nuclei.

\begin{figure}[t]
   \begin{center}
    \includegraphics[width=0.6\textwidth,clip]{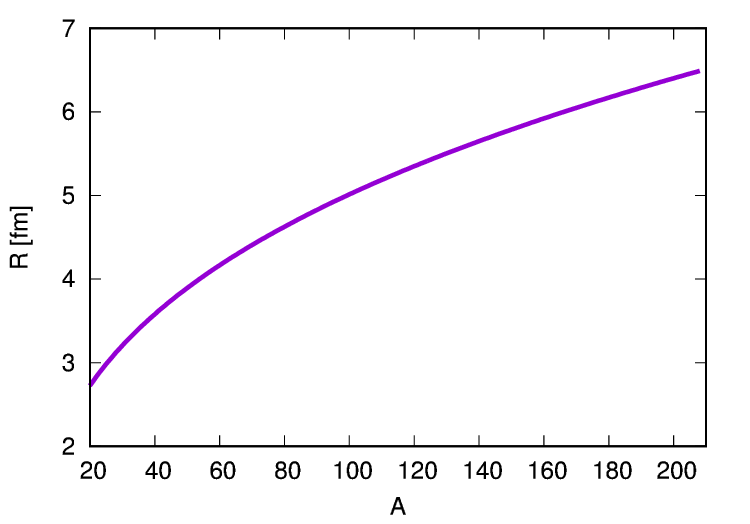}
   \end{center}
   \caption{Nucleon number ($A$) dependence of $R$ determined by
   Eq.~\eqref{R_norm}.
   }
   \label{fig:RvsA}
  \end{figure}

\begin{figure}[t]
  \begin{center}
   \includegraphics[width=0.65\textwidth,clip]{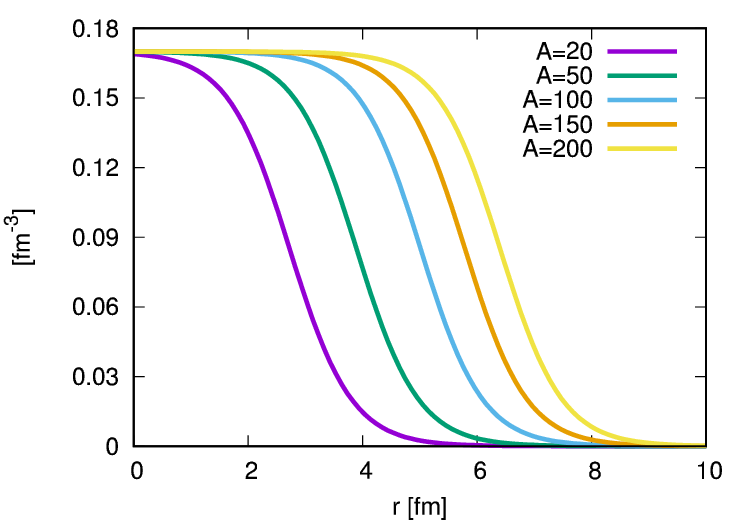}   
  \end{center}
 \caption{Nucleon number distribution functions with various nucleon numbers.
 From left to right, the distributions for $A=20,50,100,150,$ and $200$ are
 shown.
 }
 \label{fig:Nuclear_distribution}
\end{figure}

 For the atomic nuclei with $A\gtrsim16$ ($A$ is the nucleon number), the nucleon number distribution function is approximately
 expressed by the Woods-Saxon function as
 \begin{align}
  \rho(r) = \frac{\rho_0}{1+\exp\left((r-R)/a\right)} \, ,
  \label{eq:nucldist}
 \end{align}
 where $\rho_0 = 0.17$ ${\rm fm}^{-3}$ and $a= 0.54$ ${\rm
 fm}$~\cite{Mottelson196901}\footnote{For nuclei with $A<16$,
 other shapes of the distribution have been applied, such
 as a Gaussian function.}.
$r$ is the distance from the center of the nucleus, and
$R$ is determined to satisfy
\begin{align}
 \int \rho(r) d^3r=A \, . \label{R_norm}
\end{align}
$R$ as a function of $A=20,\dots,208$ is shown in Fig.~\ref{fig:RvsA}.
The nucleon number distribution functions for several $A$
 are displayed in Fig.~\ref{fig:Nuclear_distribution}.

\begin{figure}[tbp]
 \begin{center}
   \begin{tabular}{cc}
    (i) $\bar{D}$-nucleus potential& (ii) $B$-nucleus potential \\
   \includegraphics[width=0.5\textwidth,clip]{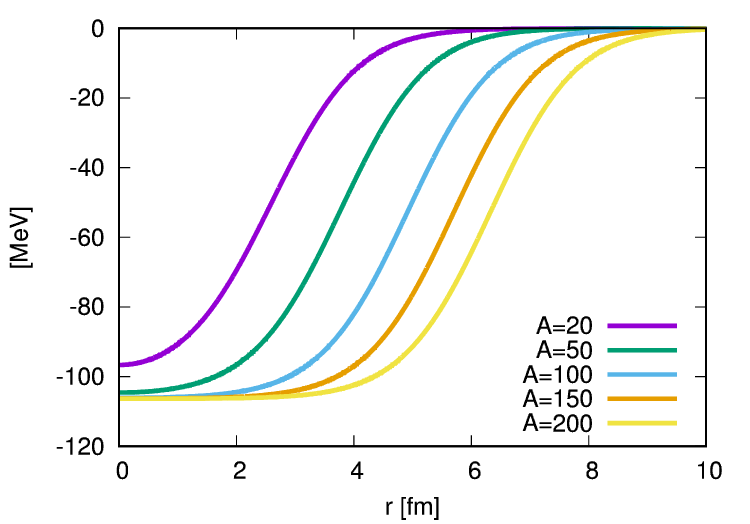} &
   \includegraphics[width=0.5\textwidth,clip]{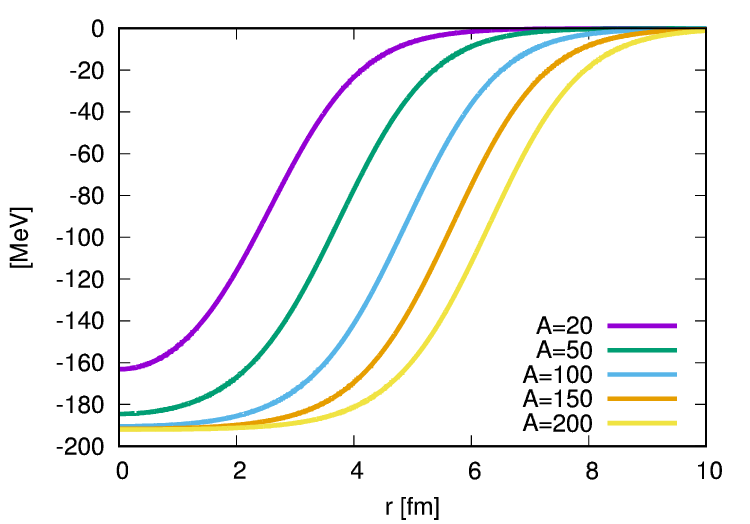}  \\
   \end{tabular}
  \caption{The folding potentials for the (i) $\bar{D}$-nucleus and (ii)
  $B$-nucleus systems.
  From left to right, the potentials for $A=20,50,100,150$ and $200$ are
  shown.}
  \label{fig:FoldingDbarNBN}
 \end{center}
\end{figure}

  From the
  single-channel potential in Eq.~\eqref{v_pro} and the nucleon number 
 distribution function in Eq.~\eqref{eq:nucldist},
 the folding potential between the $P$ meson and the nucleus is obtained by
 \begin{align}
  V_{{\rm fold}}(r)
  =\int
  V_{\rm pro}(r-r^\prime)
  \rho(r^\prime)d^3r^\prime \, .
  \label{eq:foldingpote}
 \end{align}
The obtained folding potentials of the $\bar{D}$ ($B$) meson 
 are shown in Fig.~\ref{fig:FoldingDbarNBN}.

In this study, we employ the folding potential, which is interpreted as the Hartree potential with the local density approximation.
In Eq.~\eqref{eq:foldingpote},
$V_{\rm pro}$ corresponds to the Born term of the full $t$-matrix of the $\bar{D}N$ scattering.
In the literature, as the standard approach to obtain the optical potential, 
the $t\rho$ approximation has been used~\cite{Ericson:1988gk,Hayano:2008vn}.
If the $t$-matrix has any pole, however, the breaking down of the $t\rho$ approximation can occur, 
as was presented in detail 
in Ref.~\cite{GarciaRecio:2011xt}.

Let us make a comment on the $I(J^P)=1(1/2^-)$ channel.
The interaction of this channel is repulsive as we obtained a repulsive scattering length, 
$a_{I=1}=-0.07$ fm, in Ref.~\cite{Hosaka:2016ypm}.
Based on this value, applying the $t\rho$ approximation in Eq.~(4.1.24) in Ref.~\cite{Hosaka:2016ypm}, we estimate the mass shift (attraction or repulsion) at normal nuclear matter density, and obtain a positive mass shift by 4 MeV.
However, this mass shift is smaller than the potential depth of about 100 MeV (potential value at $r=0$) for $\bar{D}N$ in $I=0$ in
Fig.~\ref{fig:FoldingDbarNBN}.
Therefore, we may approximately neglect the contribution from $I=1$ in the present research.

\section{Numerical results}
\label{sec:result}
From the folding potential in Eq.~\eqref{eq:foldingpote}, 
the binding energies of the $P$-nucleus systems are obtained by solving the 
Schr\"odinger equation for the $P$ meson and the nucleus,
\begin{align}
 -\frac{1}{2\mu}\left(\frac{\partial^2}{\partial r^2}
 +\frac{2}{r}\frac{\partial}{\partial r}-\frac{\ell(\ell+1)}{r^2}\right)\psi(r)
 +V_{{\rm fold}}(r)
 \psi(r)=E\psi(r) \, ,
 \label{eq:Schrodingereq}
\end{align}
where the reduced mass $\mu$ is given by
\begin{align}
 \frac{1}{\mu}=\frac{1}{m_P}+\frac{1}{Am_N} \, ,
 \label{eq:redeuced_mass}
\end{align}
with the $P$ meson (nucleon) mass $m_P$ ($m_N$) and the nucleon number
$A=16,\dots,208$.
$\ell$ is the relative orbital angular momentum between the $P$ meson
and the nucleus.

The Schr\"odinger equation~\eqref{eq:Schrodingereq} is solved numerically by using
the renormalized Numerov method, which was developed
in Refs.~\cite{Johnson:1977nm,Johnson:1978nm}.
We investigate not only a bound state but also a resonance state from a phase shift $\delta$.
The resonance energy $E_{\rm re}$ is determined by an inflection point of
$\delta$~\cite{Arai:1999pg,Yamaguchi:2011qw}, and
the decay width is given by $\Gamma=2/(d\delta/dE)_{E=E_{\rm re}}$.

\begin{figure}[t]
 \begin{center}
  \includegraphics[width=0.40\textwidth,clip]{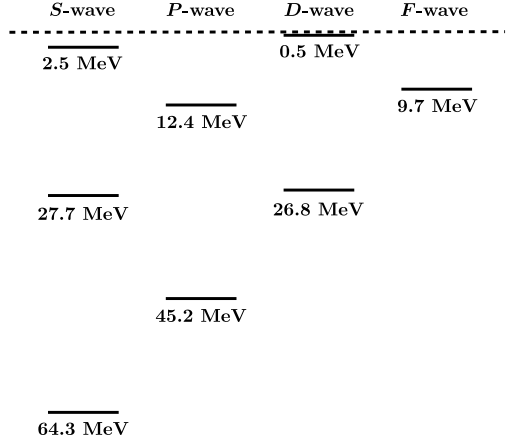}
  \caption{\label{fig:Binding_resultDbarA_S_A16}
  Energy levels of $\bar{D}-$nucleus systems with $S$, $P$,
  $D$, and $F-$waves for $A=16$.
  The solid lines are the energy levels obtained.
  The values are the binding energies.
  The dashed line is the threshold.
  }  
 \end{center} 
\end{figure}

\begin{figure}[t]
    \begin{center}     
    \begin{tabular}{cc}
     (i) $S$-wave & (ii) $P$-wave\\
     \includegraphics[width=0.45\textwidth,clip]{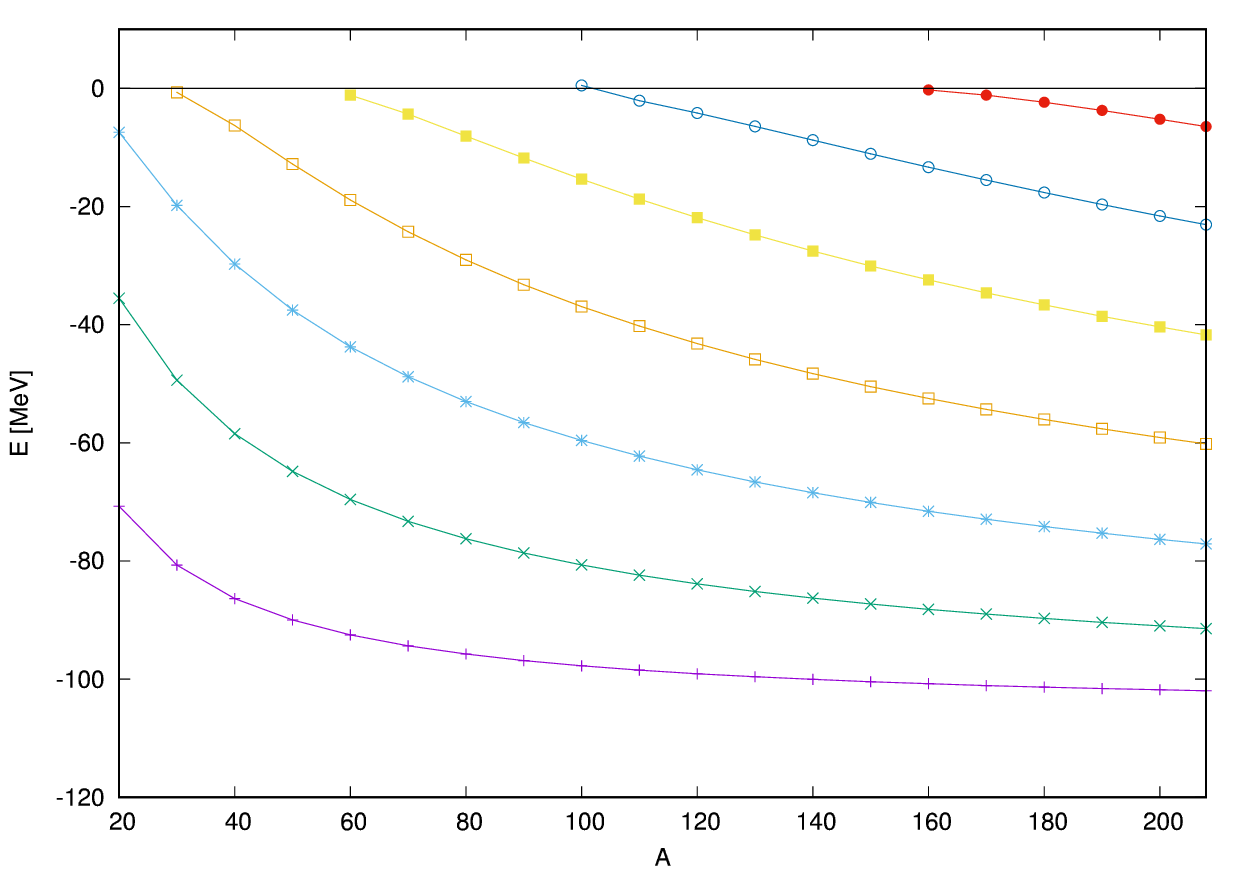}  &
	 \includegraphics[width=0.45\textwidth,clip]{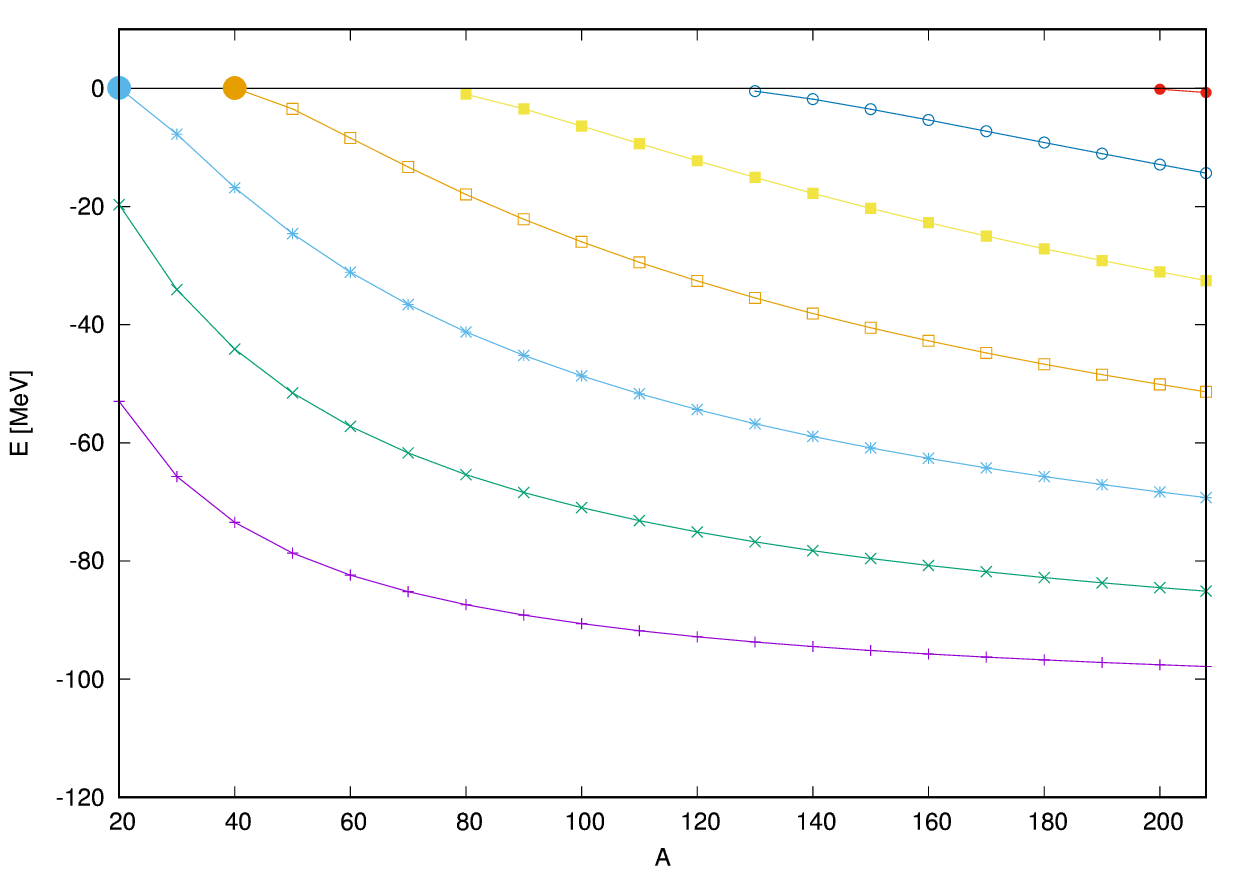}  	 \\
     (iii) $D$-wave & (iv) $F$-wave\\
        \includegraphics[width=0.45\textwidth,clip]{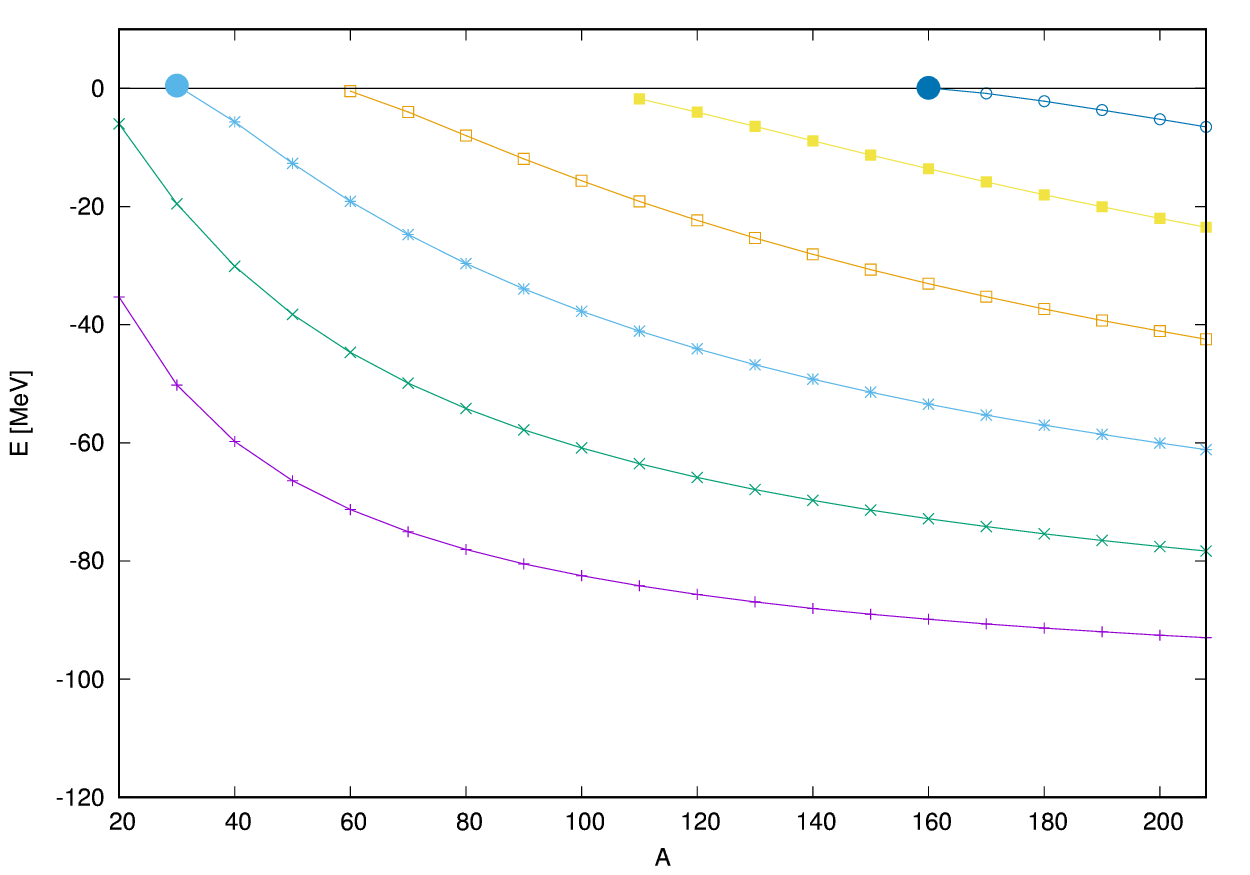}     &
	 \includegraphics[width=0.45\textwidth,clip]{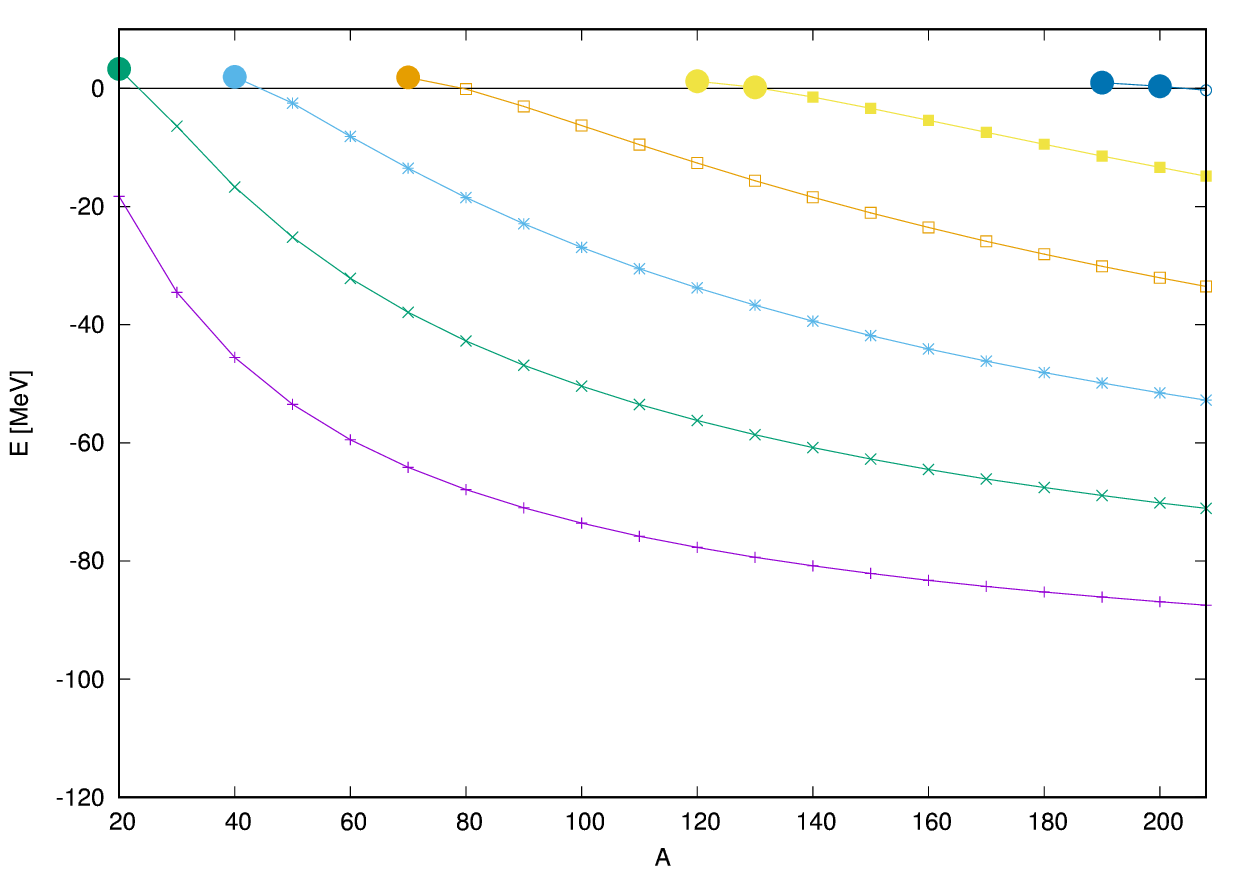}  	 \\
    \end{tabular}
     \caption{Nucleon number $A$ dependence on the binding energies of the
     $\bar{D}$-nucleus systems with (i) $S$-wave, (ii) $P$-wave, (iii) $D$-wave, and
     (iv) $F$-wave.
     The large filled circle shows resonances.
     The energy is measured from the $\bar{D}$-nucleus threshold, and given
     in units of MeV.}
     \label{fig:Binding_resultDbarA_S}
    \end{center}
   \end{figure}

\begin{figure}[t]
 \begin{center}
  \includegraphics[width=0.40\textwidth,clip]{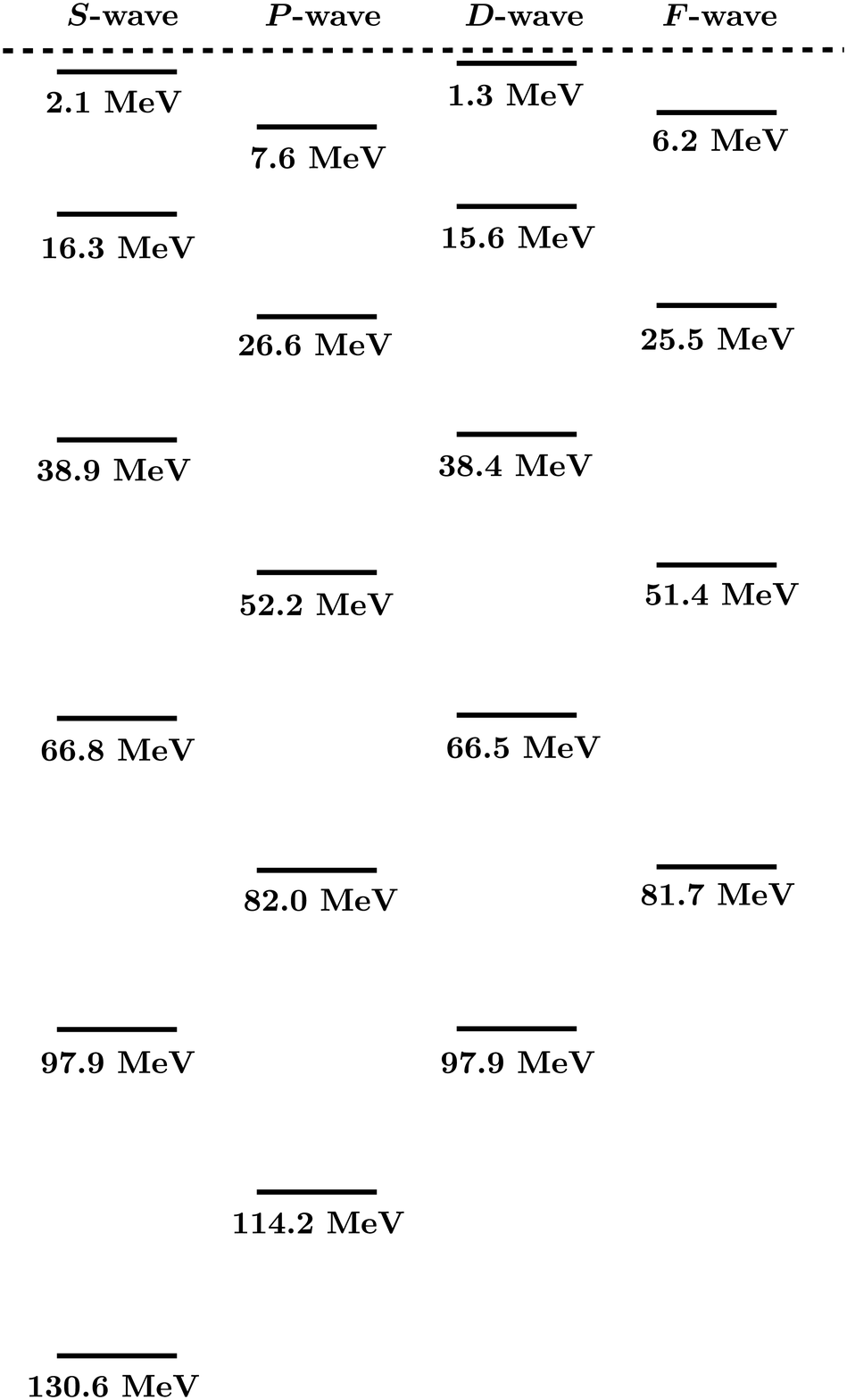}
  \caption{\label{fig:Binding_resultBA_S_A16}
  Energy levels of $B-$nucleus systems with $S$, $P$,
  $D$, and $F-$waves for $A=16$.
  The solid lines 
  are the energy levels obtained.
  The values are 
  the binding energies.
  The dashed line is the threshold.
  }  
 \end{center} 
\end{figure}

\begin{figure}[t]
    \begin{center}     
    \begin{tabular}{cc}
     (i) $S$-wave & (ii) $P$-wave\\
     \includegraphics[width=0.45\textwidth,clip]{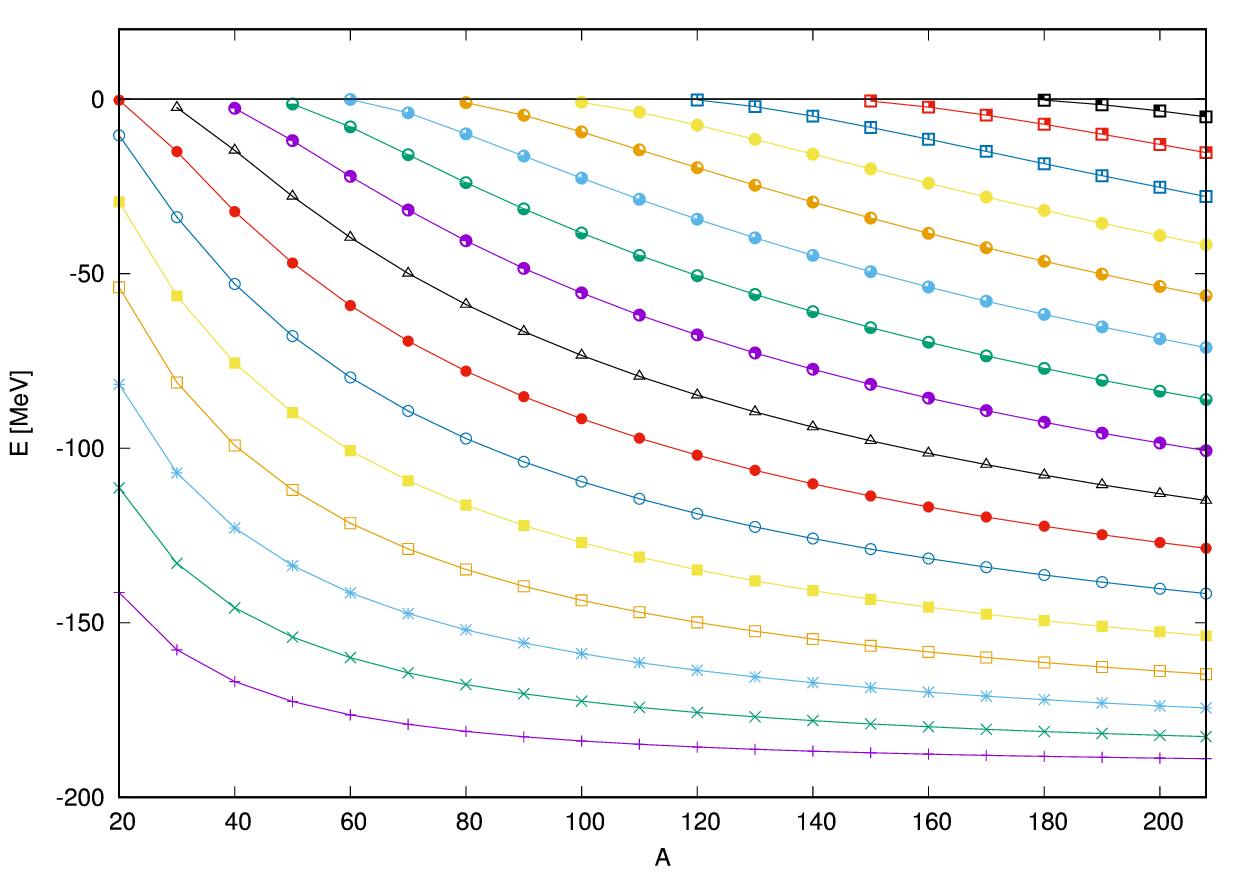}  &
	 \includegraphics[width=0.45\textwidth,clip]{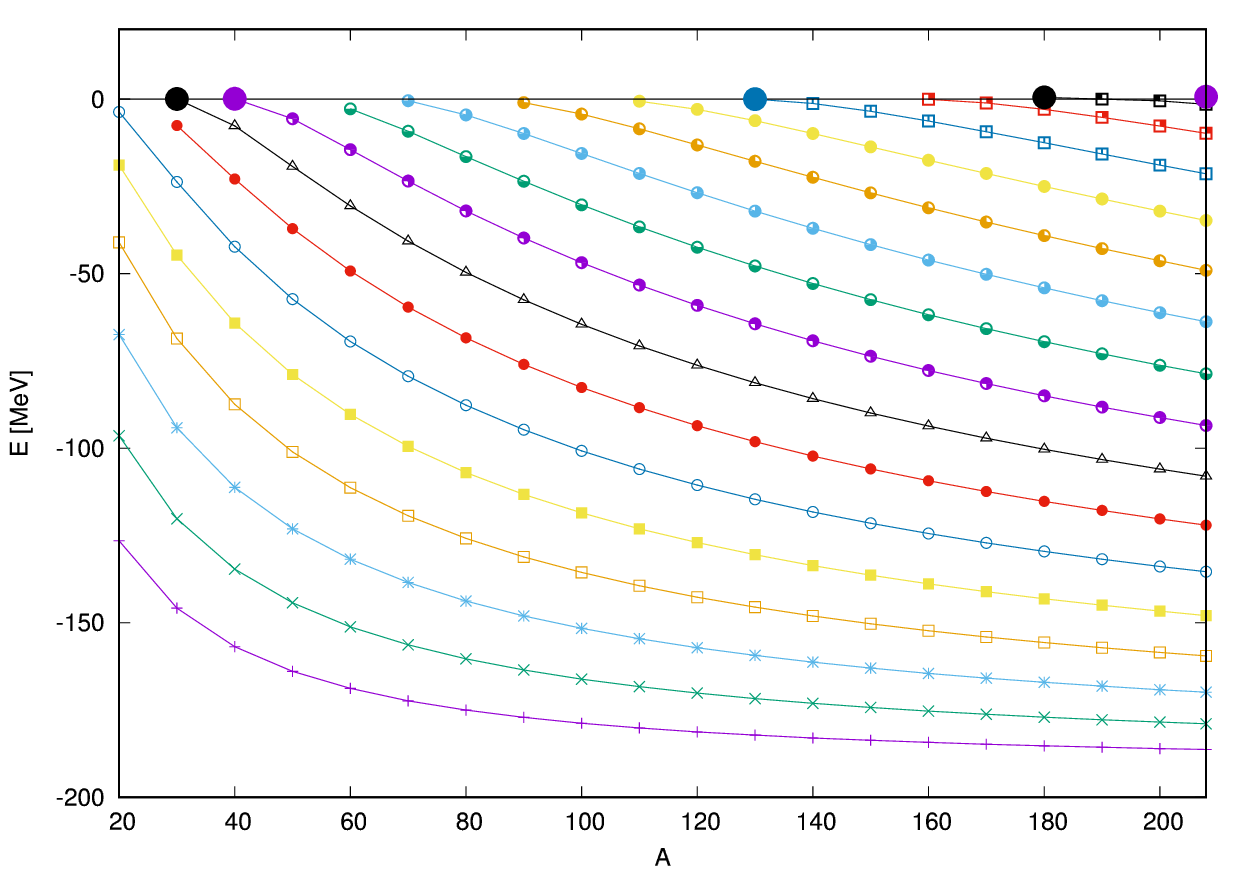}  	 \\
     (iii) $D$-wave & (iv) $F$-wave\\
        \includegraphics[width=0.45\textwidth,clip]{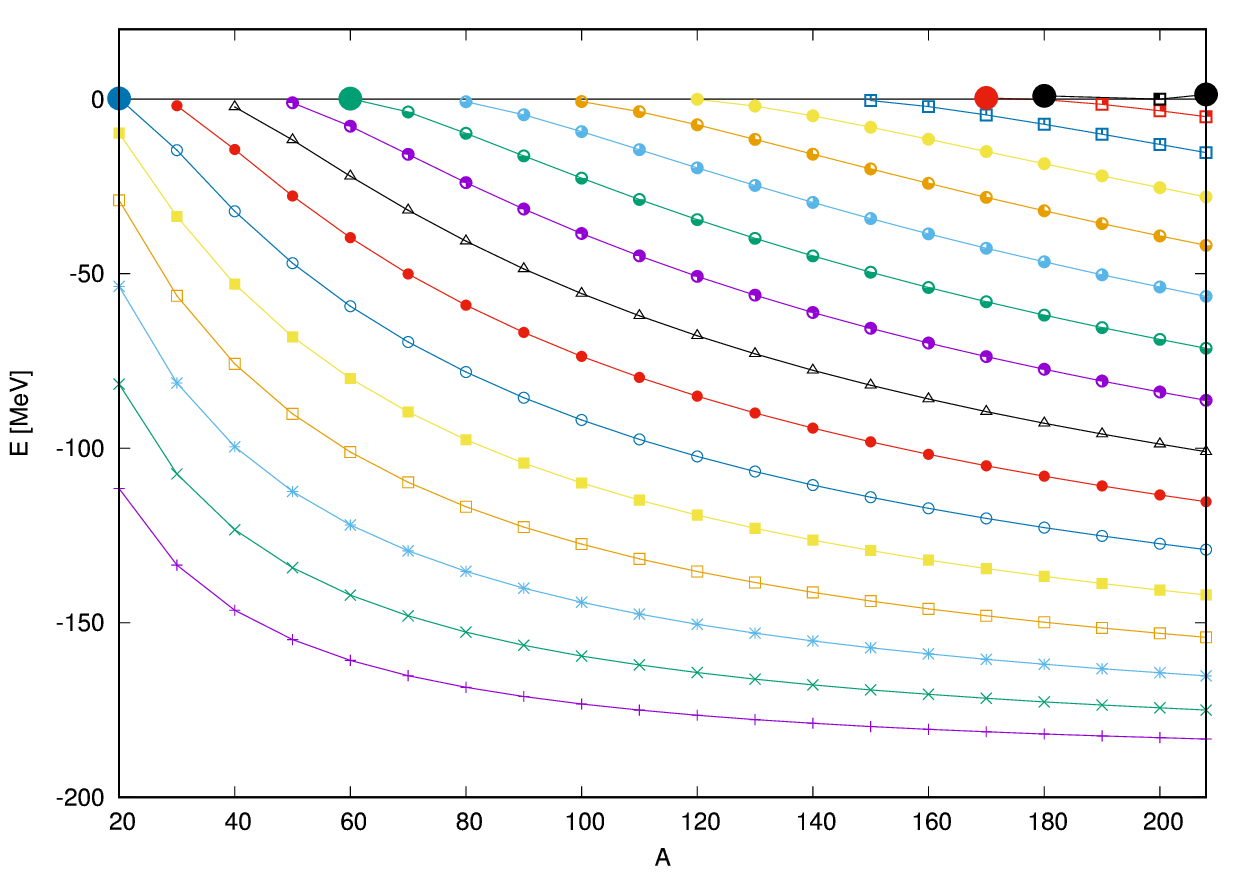}     &
	 \includegraphics[width=0.45\textwidth,clip]{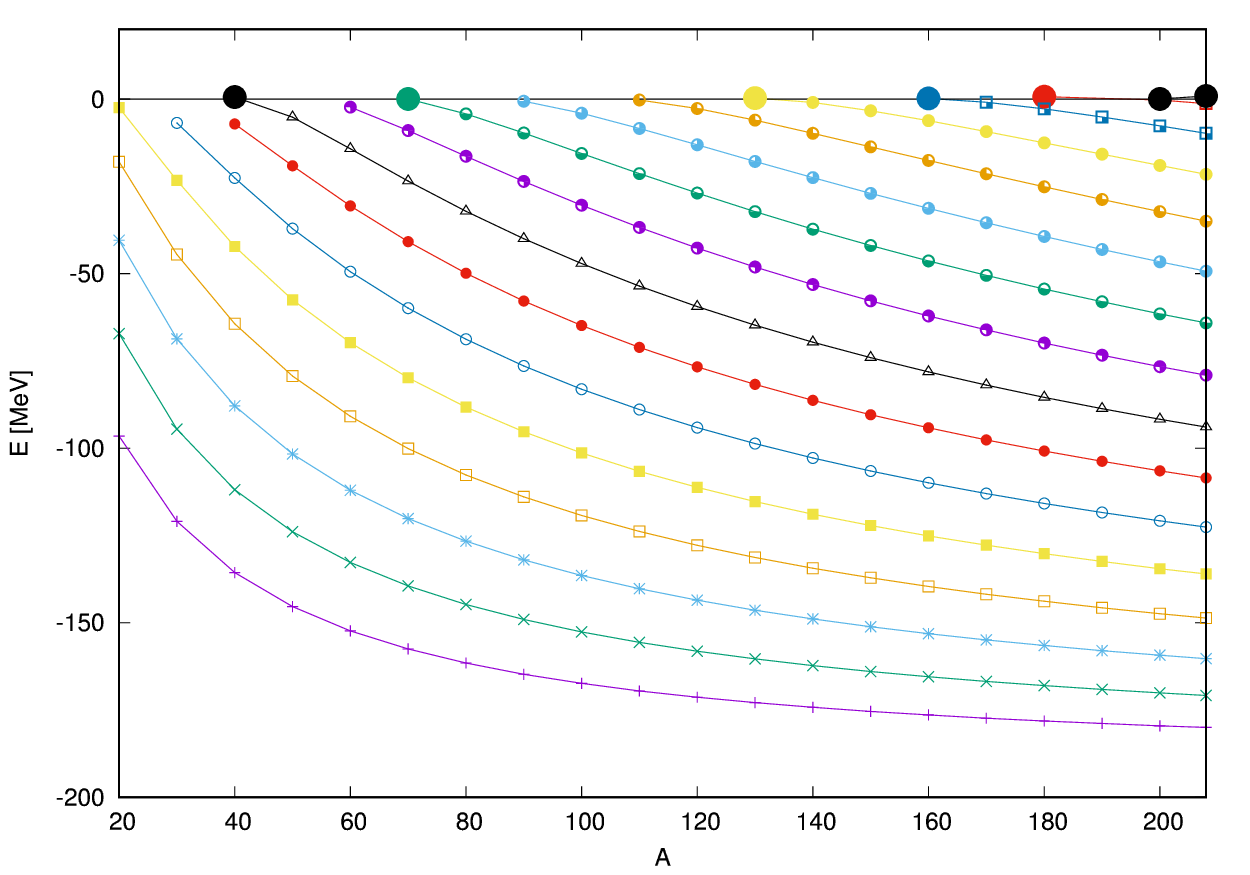}  	 \\
    \end{tabular}
     \caption{Nucleon number $A$ dependence 
     on binding energies of the
     $B$-nucleus systems with (i) $S$-wave, (ii) $P$-wave, (iii) $D$-wave, and (iv)
     $F$-wave.
     Same convention as Fig.~\ref{fig:Binding_resultDbarA_S}.
     }
     \label{fig:Binding_resultBA}
    \end{center}
   \end{figure}

The energies are computed for $\ell=0,1,2,3$, namely $S$, $P$, $D$, and
$F$-waves.
For the charm sector,
the energies obtained are shown in
Fig.~\ref{fig:Binding_resultDbarA_S_A16} for $A=16$
which is the minimal $A$ to 
apply Eq.~\eqref{eq:nucldist}, and Fig.~\ref{fig:Binding_resultDbarA_S}
with various nucleon numbers for $A=20,\dots,208$.
The energies of the bottom sector are shown in
Figs.~\ref{fig:Binding_resultBA_S_A16}
and \ref{fig:Binding_resultBA} for $A=16$ and $A=20,\dots,208$, respectively.

\begin{table}[htbp]
 \caption{Resonance energy $E_{\rm re}$ and half-decay width $\Gamma/2$
 for the $\bar{D}$-nucleus systems with $P$, $D$, and $F$-waves.
 $E_{\rm re}$ and $\Gamma/2$ are given in units of MeV.}
 \label{table:resonance_Dbar}
   \begin{center}
     \begin{tabular}{lp{3em}p{7em}c}
      \hline\hline
     $\ell$&$A$ &$E_{\rm re}$ [MeV] &$\Gamma/2$ [MeV] \\
      \hline
      $P$-wave &20 &$8.90\times 10^{-2}$&$3.47\times 10^{-2}$ \\
      &40 &$5.40\times 10^{-2}$&$2.02\times 10^{-2}$ \\ \hline
      $D$-wave&30 &$4.82\times 10^{-1}$&$7.20\times 10^{-2}$ \\
      &160 &$1.05\times 10^{-1}$&$8.40\times 10^{-3}$ \\ \hline
      $F$-wave&20 &$3.29$&$8.09\times 10^{-1}$ \\
      &40 &$1.95$&$4.35\times 10^{-1}$ \\
      &70 &$1.85$&$7.53\times 10^{-1}$ \\
      &120 &$1.23$&$4.13\times 10^{-1}$ \\
      &130 &$1.85\times 10 ^{-1}$&$1.42\times 10^{-3}$ \\
      &190 &$9.77\times 10^{-1}$&$4.46\times 10^{-1}$ \\
      &200 &$3.59\times 10^{-1}$&$2.71\times 10^{-2}$ \\
      \hline\hline
     \end{tabular}
   \end{center}
\end{table}

For the charm sector,
we find bound states for 
the $S$, $P$, $D$, and $F$-wave states,
as shown in
Figs.~\ref{fig:Binding_resultDbarA_S_A16} and \ref{fig:Binding_resultDbarA_S}.
We emphasize that these bound states are stable against the strong decay.
For the $S$-wave state,
three bound states (ground and the first and second excited states)
are obtained
in $A=16$.
As seen in Fig.~\ref{fig:Binding_resultDbarA_S},
the number of 
bound states increases when the nucleon number increases.
Finally, seven bound states are obtained in $A=208$.
The binding energy also increases as $A$ increases.
However,
the binding energies with respect to $A$ become saturated in the
large-$A$ region, because
the nucleon number distribution also becomes saturated and
the $A$ dependence of the reduced mass (\ref{eq:redeuced_mass}) becomes smaller.
Therefore, the $A$ dependence on 
the binding energies is flat in the large-$A$ region.

 For the states with $\ell\neq 0$,
 the number of 
 bound states and the values of the binding energy
 are reduced in comparison with 
 those of $\ell=0$.
 As a new phenomenon, resonant states appear slightly above the thresholds.
 The properties of the resonances are summarized in
 Table~\ref{table:resonance_Dbar}.
 Interestingly, 
 sharp resonances with 
 small widths less than 1 MeV are found.
 In particular, the $F$-wave states have many resonances.

\begin{table}[htbp]
 \caption{Resonance energy $E_{\rm re}$ and half-decay width $\Gamma/2$
 for the $B$-nucleus systems with $P$, $D$, and $F$-waves.
 Same convention as Table~\ref{table:resonance_Dbar}. }
  \label{table:resonance_B}
   \begin{center}
     \begin{tabular}{lp{3em}p{7em}c}
      \hline\hline
      $\ell$&$A$ &$E_{\rm re}$ [MeV] &$\Gamma/2$ [MeV] \\
      \hline
      $P$-wave
      &30 &$4.90\times 10^{-2}$&$5.10\times 10^{-2}$ \\
      &40 &$9.30\times 10^{-3}$&$2.94\times 10^{-3}$ \\
      &130 &$1.33\times 10^{-2}$&$8.00\times 10^{-3}$ \\
      &180 &$4.84\times 10^{-1}$&$4.46\times 10^{-4}$ \\
      &208 &$6.94\times 10^{-1}$&$1.60\times 10^{-4}$ \\
      \hline
      $D$-wave
      &20 &$2.27\times 10^{-1}$&$5.95\times 10^{-2}$ \\
      &60 &$1.38\times 10^{-1}$&$4.58\times 10^{-2}$ \\
      &170 &$3.42\times 10^{-1}$&$1.25\times 10^{-2}$ \\
      &180 &$9.30\times 10^{-1}$&$5.70\times 10^{-3}$ \\
      &208 &$1.36$&$3.60\times 10^{-3}$ \\
      \hline
      $F$-wave
      &40 &$5.94\times 10^{-1}$&$2.76\times 10^{-1}$ \\
      &70 &$3.47\times 10^{-2}$&$6.05\times 10^{-5}$ \\
      &130 &$3.21\times 10^{-1}$&$1.36\times 10^{-1}$ \\
      &160 &$1.91\times 10^{-1}$&$3.76\times 10^{-2}$ \\
      &180 &$7.12\times 10 ^{-1}$&$9.25\times 10^{-4}$ \\
      &200 &$5.96\times 10^{-2}$&$1.67\times 10^{-2}$ \\
      &208 &$8.97\times 10^{-1}$&$2.72\times 10^{-4}$ \\
      \hline\hline
     \end{tabular}
   \end{center}
\end{table}

For the bottom sector,
the attraction of the $B$-nucleus potential is stronger than the
$\bar{D}$-nucleus one.
This is because the mass difference between $B$ and $B^{\ast}$ is smaller in the original potential and the strong mixing between $BN$ and $B^{\ast}N$ channels induces the stronger attraction~\cite{Yasui:2009bz,Yamaguchi:2011xb,Yamaguchi:2011qw}.
Therefore, more bound states are found.
For the $S$-wave state,
the number of 
bound states is
six in $A=16$ in Fig.~\ref{fig:Binding_resultBA_S_A16}.
In $A=208$, sixteen states appear
as shown in Fig.~\ref{fig:Binding_resultBA},
and the binding energy of the ground
state reaches $188.9$ MeV.
The behavior of the binding energy with respect to $A$
is similar 
to the $\bar{D}$-nucleus systems.
For $\ell\neq 0$, we find the resonant states 
that 
are summarized in Table~\ref{table:resonance_B}.
The $B$-nucleus systems produce many resonances with 
small widths.

      \begin{figure}[t]
     \begin{center}
    \begin{tabular}{ccc}
	 \includegraphics[width=0.3\textwidth,clip]{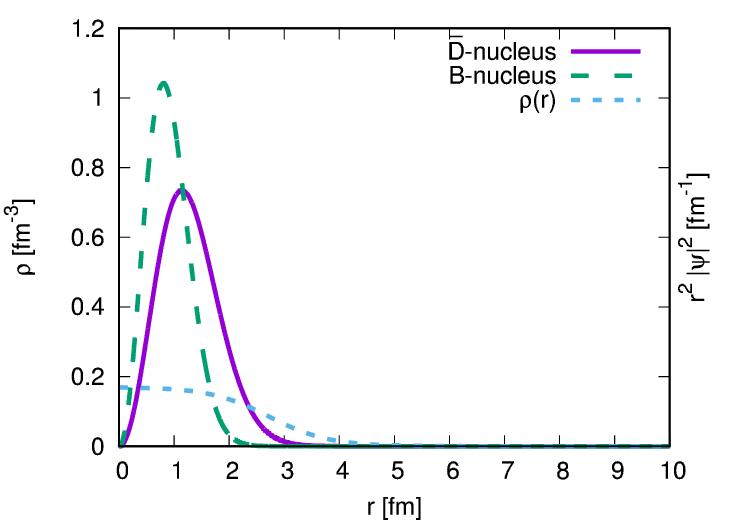}  &
	 \includegraphics[width=0.3\textwidth,clip]{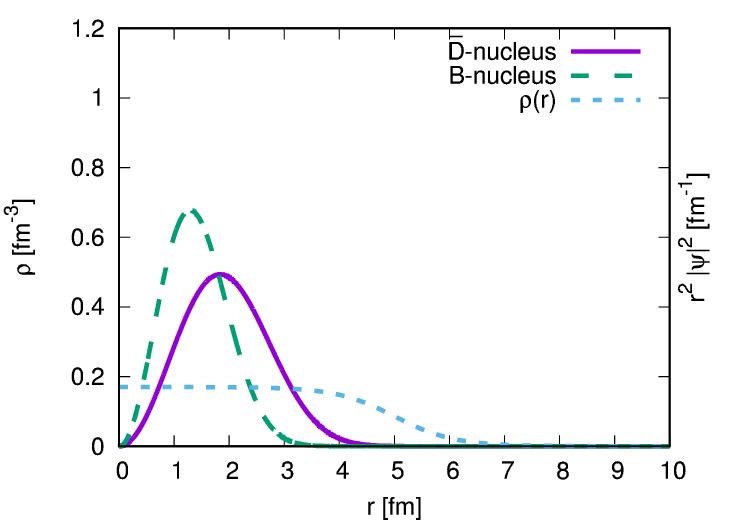}     &
	     \includegraphics[width=0.3\textwidth,clip]{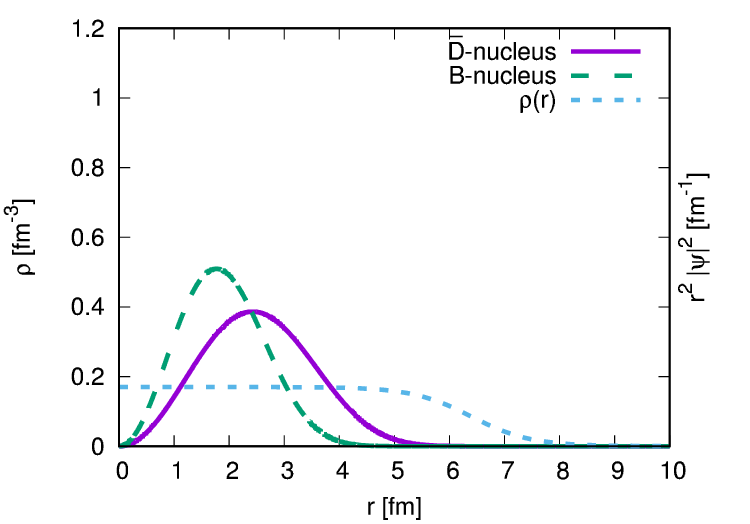}   \\
    \end{tabular}
      \caption{The wave functions of the ground states with $S$-wave,
      $r\psi(r)$,
      and the nucleon number
      distribution function, $\rho (r)$, for $A=20$ (left), 
      $A=100$ (middle), and $A=200$ (right).
      The solid (dashed) lines show
      $r^{2}|\psi(r)|^{2}$ for the
     $\bar{D}$-nucleus (the $B$-nucleus), and
     the dotted lines show $\rho(r)$.
     }
     \label{fig:Wave_func}
     \end{center}
   \end{figure}

The wave functions obtained for the $\bar{D}$, $B$ mesons and the nucleon
number distribution function $\rho(r)$ are plotted for $A=20,100,200$ in
Fig.~\ref{fig:Wave_func}.
The wave functions of the $\bar{D}$, $B$ mesons are localized inside the nuclei.
The wave function of the $B$-nucleus shrinks 
more 
than that 
of the $\bar{D}$-nucleus because of the stronger attraction for the former.
As the nucleon number increases, the radii of the nuclei become larger, and accordingly  the wave functions of the $\bar{D}$, $B$ mesons become extended.

For 
discussion, we compare our results with 
those 
in Ref.~\cite{GarciaRecio:2011xt}.
First of all, we notice that the interaction between a $\bar{D}$-meson
and a nucleon in Ref.~\cite{GarciaRecio:2011xt}
was obtained 
with a vector-meson exchange model with flavor-spin SU(8) symmetry.
The exchanged mesons as well as the method for including the finite-size effect are different from ours.
Nevertheless, both results show some similarities.
In both cases, there are several excited states from the ground state, and higher angular momentum states are 
able 
to exist.
As a slight difference, the $\bar{D}^{0}$ meson energies in nuclei are saturated at nucleon number $A=40$ in Ref.~\cite{GarciaRecio:2011xt}, while they become saturated for $A \gtrsim 100$ in our analysis.
 We emphasize that, as a new phenomenon in the present analysis, there exist several resonant states generated by the centrifugal barrier potential near thresholds.

We also comment that the wave function of the $\bar{D}$ meson 
in $^{208}\mathrm{Pb}$ was obtained in Ref.~\cite{Tsushima:1998ru}.
Though the interaction used is different from ours, the wave functions obtained are comparable with ours.

\section{Summary}
   We have studied the bound and resonant systems of the heavy meson ($P=\bar{D}$ or
   $B$) and the nucleus with nucleon number
   $A=16,\cdots,208$.
   The attraction between the $P(=\bar{D},B)$ meson and nucleon $N$,
   which is enhanced by the $PN-P^\ast N$ mixing, inspires us to investigate
   the stability of exotic heavy mesic nuclei 
   against the strong decay.
   The mesic nuclei are analyzed as 
   two-body systems of the $P$ meson
   and the nucleus, where the interaction is described by the folding
   potential.
   By using the single-channel $PN$ potential and the nucleon number
   distribution function,
   the folding potential is obtained.

   We solve the Schr\"odinger equations of the two-body $P$-nucleus
   system with the nucleon number
   $A=16,\dots,208$.
   Many bound and resonant states are obtained as a result.   
   We find that the binding energy increases as the nucleon number $A$
   increases.
   The $A$ dependence on 
   the binding energy becomes flat in the
   large-$A$ region.
   
   In the present research, we do not include the possible short-range core in the interaction between a $\bar{D}$ ($B$) meson and a nucleon.
   We do not consider the Coulomb potential, which can be important for $D^{-}$ ($B^{+}$) mesons in large nuclei, 
   as discussed in Refs.~\cite{Tsushima:1998ru,GarciaRecio:2011xt}. 
   It would also be 
   interesting to investigate the $D_{s}^{-}$ ($B_{s}^{0}$) mesons in atomic nuclei.
    Further theoretical discussions are left for future work.
    The information on the energy spectra of 
   open-heavy mesic nuclear systems will be useful
   for 
   future experimental research 
   at 
   the Facility for Antiproton and Ion Research (FAIR), 
   the Japan Proton Accelerator Research Complex (J-PARC),
   the Relativistic Heavy Ion Collider (RHIC), 
   the Large Hadron Collider (LHC), and so forth.

   \section*{Acknowledgments}
    This work is supported in part by the Istituto Nazionale di Fisica
    Nucleare (INFN) Fellowship Programme and
    the Special Postdoctoral Researcher (SPDR) Program of RIKEN (Y.Y.).
    This work is also supported by 
    Grants-in-Aid for Scientific Research
    (Grant No.~25247036 and No.~15K17641) from the Japan Society for the
    Promotion of Science (JSPS) (S. Y.).


%

\end{document}